\documentclass{pasj00}

\begin{document}
\SetRunningHead{M. Sasada et al.}{Microvariability of S5~0716$+$714}
\Received{2008/7/19}
\Accepted{2008/10/10}

\title{Detection of Polarimetric Variations Associated with the Shortest
Time-Scale Variability in S5~0716$+$714}

\author{
Mahito \textsc{Sasada}\altaffilmark{1},
Makoto \textsc{Uemura}\altaffilmark{2},
Akira \textsc{Arai}\altaffilmark{1},
Yasushi \textsc{Fukazawa}\altaffilmark{1},\\
Koji S. \textsc{Kawabata}\altaffilmark{2},
Takashi \textsc{Ohsugi}\altaffilmark{2},
Takuya \textsc{Yamashita}\altaffilmark{2},
Mizuki \textsc{Isogai}\altaffilmark{2},
Shuji \textsc{Sato}\altaffilmark{3},
and Masaru \textsc{Kino}\altaffilmark{3}}

\altaffiltext{1}{Department of Physical Science, Hiroshima University, Kagamiyama 1-3-1, Higashi-Hiroshima 739-8526}
\email{sasada@hep01.hepl.hiroshima-u.ac.jp}
\altaffiltext{2}{Astrophysical Science Center, Hiroshima
University, Kagamiyama 1-3-1, Higashi-Hiroshima 739-8526}
\altaffiltext{3}{Department of Physics, Nagoya University, Furo-cho, Chikusa-ku, Nagoya 464-8602}


%


\maketitle

\begin{abstract}
We present the result of near-infrared and optical
 observations of the BL Lac object S5~0716$+$714 carried out by the KANATA
 telescope. S5~0716$+$714 has both a long term high-amplitude
 variability and a short-term variability within a night. The shortest
 variability (microvariability)
 time-scale is important for understanding the geometry of jets and
 magnetic field, because it provides a possible minimum size of
 variation sources. Here, we report the detection of 15-min variability
 in S5~0716$+$714, which is one of the shortest time-scales in
 optical and near-infrared variations observed in blazars. The detected
 microvariation had an amplitude of $0.061{\pm}0.005$~mag in $V$
 band and a blue color of $\Delta(V-J)=-0.025{\pm}0.011$. 
 Furthermore, we successfully detected an unprecedented, short
 time-scale polarimetric variation which correlated with the brightness
 change. We revealed that the microvariation
 had a specific polarization component. The polarization degree of the
 variation component was
 higher than that of the total flux. These results suggest that the
 microvariability originated from a small and local region where the
 magnetic field is aligned.
\end{abstract}

\section{Introduction}
Blazars are one of classes of active galactic nuclei (AGNs), whose
relativistic jets are considered to be directed along the line of sight
(Blandford \& K\"onigl, 1979 , \cite{Antonucci}).
Blazars show a large and rapid variability, and their fluxes
often vary by an order of magnitude in a time-scale of days. A short
time-scale variability has also been reported, that is, so-called
intra-day variability or microvariability, having a time-scale of
$<1$~day (e.g. \cite{Antonucci}).

The observation of the shortest time-scale variability gives us important
information for understanding the variation mechanism, the emitting
region and the jet geometry. An extremely large variability
with a time-scale of tens of minutes has recently been reported in the
optical region (e.g. \cite{Xie} for eight blazars e.g. S5~0716$+$714). However,
\citet{Cellone07a} argued that
they can be spurious results due to systematic errors arising from analysis.
The shortest time-scale in blazar variability was, in general,
believed to be several hours.

In gamma-ray region, however, a significant variability with a
time-scale of minutes have recently been discovered by \citet{Albert} in
Mrk~501 and \citet{Aharonian} in PKS2155$-$304.
Also in the optical band, a 15-min microvariation has
been reported by \citet{Gupta} in Mrk~501. The amplitude of the
microvariation was not so extremely large, but rather small, with
0.05~mag in $R$ band.
Such a short time-scale indicates that the size of the emitting region of
the variation is smaller than a Schwarzschild radius $R_{\rm S}$ of
the central black hole in AGN if we assume typical values for the black
hole mass and the jet
Doppler factor. It is, hence, possible that microvariations originate
from a small confined region rather than a whole emitting area. On the
other hand, observational features of those shortest time-scale
variations have not been established due to the lack of detailed
observations, for example, polarimetric and multi-wavelength observations.

The spectral energy distribution (SED) of blazars can be described by
synchrotron radiation and inverse-Compton scattering. In the optical
band, the synchrotron radiation is dominant. Variations of
polarization can, hence, be observed in blazars (e.g. Angel \& Stockman,
1980). 
The polarimetric observation provides an important clue for the magnetic
field and its geometry. However, there is still few polarimetric
observation of microvariability so far. 
\citet{Cellone07b} observed hours-scale variations in the polarization
of AO~0235$+$164. According
to their observation, the temporal variation of polarization parameters
showed no correlation with the variation of the total flux.
For the shortest time-scale variations (${\sim} 10$~min), there was no
report studying possible correlation between the polarization and the
total flux.

S5~0716$+$714 is a BL Lac object, and one of systems in which
microvariability was observed. The variability has been observed on
the time-scale of days and hours (e.g. \cite{Wagner}, \cite{Stalin06}
and \cite{Wu07}). \citet{Nilsson08}
estimated its redshift to be $z = 0.31 \pm 0.08$.

In order to study the mechanism of the microvariability and to find the
shortest time-scale variations, we performed photopolarimetric
observations of S5~0716$+$714 in the optical and
near-infrared (NIR) bands simultaneously. In this paper, we report our
detection of 15-min variability with polarimetric variations. The 15-min
variability has a small amplitude, but the light-curve is 
very well-sampled. The paper is arranged as follows: In section 2,
we present the observation method and analysis. In section 3, we report
the result of the photometry and polarimetry. We discuss the emitting
region and the geometry of the magnetic field based on the observation
in section 4. Finally, the conclusion is drawn in section 5.

\section{Observation}
In this paper, we focus on the observation on 20 Oct 2007. It is part
of our monitoring of this blazar in 2007 using TRISPEC attached to the
1.5-m ``KANATA'' telescope at Higashi-Hiroshima
Observatory. TRISPEC has a CCD and two InSb arrays, producing
photopolarimetric data in an optical band and two NIR bands
simultaneously (\cite{Watanabe}).
A unit of the observing sequence consisted of successive exposures at
four position angles of the half-wave plate; 0\arcdeg, 45\arcdeg,
22\arcdeg 5, and 67\arcdeg 5.

The integration times in each exposure were 63, 55 and 28 sec for $V$-,
$J$- and $K_{\rm S}$-band images, respectively.
Our one polarimetric data was derived from each set of the four
exposures. Thus, the time resolution of our polarimetry is about 5-min,
while that of our photometry is about 1-min.
The sky condition was clear and stable during our observation. The full
width of half maximum of the point spread function was typically 3 arcsec. 
All images were bias-subtracted and flat-fielded, and then we performed
aperture photometry with the {\it IRAF APPHOT} package.
We adopted differential photometry with a comparison star taken in the
same frame, and the aperture size was about 7 arcsec.
Its position is R.A.=$\timeform{07h21m52.3s}$,
Dec=$\timeform{+71D18'17.6''}$ (J2000.0), and the magnitudes are 12.48, 11.32 and
10.98 mag in $V$, $J$, and $K_{\rm S}$ bands, respectively
(\cite{Gonzalez-perez, Skrutskie}).
We confirmed that the comparison star kept constant during the
observation. The dispersion of the magnitudes of the comparison star was
less than 0.005~mag, calculated with a check star at 
R.A.$=\timeform{07h21m54.4s}$, Dec$=\timeform{+71D19'21.3''}$ which is
also taken in the same frame of S5~0716$+$714, and its magnitudes of
that are 13.55 and 12.34 in V and J bands. The comparison and the check
stars are listed in \citet{Ghisellini} as star B and C, respectively.

The instrumental polarization was confirmed to be smaller than 0.1 \%
in $V$ band by observing unpolarized standard stars. We applied no
correction. The zero point of the polarization angle is corrected as the
standard system ( measured from north to east ) by observing the polarized stars.
For NIR bands, the polarimetric accuracy was worse ($>0.4$ \%) and we
used the polarization parameter only in $V$ band in this paper.

\section{Result}
\subsection{Photometric observation}
\citet{Raiteri03} reported that the object is at about 13~mag and 15~mag
in $V$ band in its bright and faint states, respectively, based on their
8-year light curve from 1994 to 2002. In Oct 2007, the object was in a
bright state at 12.6---13.5~mag. Figure~1 shows the light curves on 20
Oct. 2007 when the object was the brightest in the period of our monitoring. 
\begin{figure}
  \begin{center}
 \FigureFile(80mm,65mm){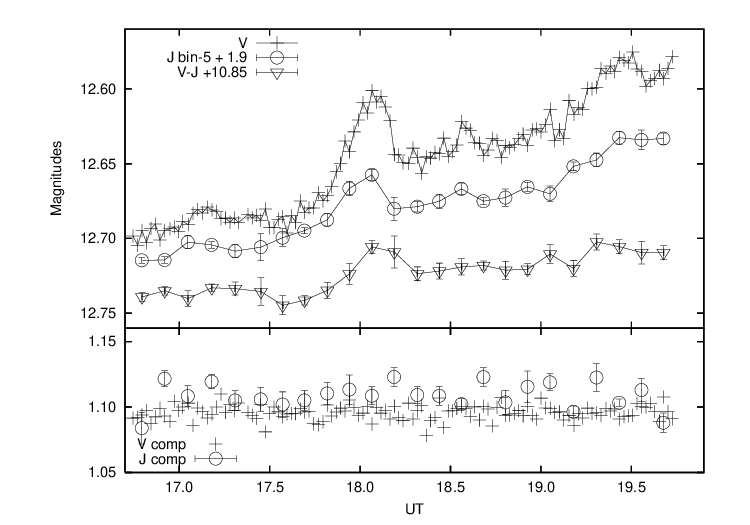}
  \end{center}
  \caption{%
  Top panel: Light curves of $V$, $J$ bands and $V-J$
 color on 20 Oct. 2007. The time bin sizes are 1- and 6-min in $V$ and
 $J$ bands, respectively. Bottom panel: Differential
 magnitudes of the comparison star against the check star in $V$ and $J$
 bands, indicating that the magnitude of the comparison star was
 constant within 0.005 and 0.010~mag in $V$ and $J$ bands, respectively. }%
  \label{fig1}
\end{figure}
During 3 hours of our observation on 20 Oct., the object brightened by
$0.123{\pm}0.005$ and $0.101{\pm}0.010$~mag in $V$ and $J$ bands,
respectively. In addition, there was a
bump-like structure around 18.0 (UT) on the overall brightening
trend. The rising time of the bump was about 15-min both in $V$ and
$J$ bands. The amplitude of the bump was $0.061{\pm}0.005$ and 
$0.035{\pm}0.010$~mag in $V$ and $J$ bands, respectively. We did not
discuss the light curve in $K_{\rm S}$ band, because the error of each
photometric point is large, about 0.1~mag.

The time-scale ${\Delta}{\tau}$ of the bump is estimated as 
\begin{eqnarray}
{\Delta}{\tau} = \frac{1}{1+z}\frac{({\Delta} F)}{dF/dt} ,
\end{eqnarray}
where $1+z$ stands for a cosmological effect (\cite{Romero}).
The rising and decaying time-scales of the bump were 970 and 620 sec in
the observer frame. These values were estimated after the
overall brightening trend was subtracted. With $z=0.31{\pm}0.08$, the
rising and decaying time-scales are calculated as $740{\pm}50$ and
$480{\pm}30$~sec in the object frame, respectively.

Cellone et~al. (2000, 2007) insisted that the
violent variations of tens of minutes time-scale may be spurious.
We now examine the 15-min variation that we detected in S5~0716$+$714
along with the points proposed by Cellone et~al. (2000, 2007).
First, the spurious variation in the light curve can arise from the variation 
of the seeing size, because of the effect of the host galaxy
(\cite{Cellone00}).
However, the host galaxy of S5~0716$+$714 is so faint that its
contribution is negligible to our photometry (\cite{Nilsson08}).
Moreover, we confirmed that there is no correlation between the light
curve and the temporal variation of the seeing size in our case.
Second, if the comparison star is $\sim$2 mag brighter
than the source, the variation is severely overestimated (\cite{Cellone07a}).
The magnitude of comparison is 12.48~mag in
$V$ band as mentioned in section~2, and that of the object was
about 12.6 in the same band. The difference of magnitude in the
object and the comparison is $< 1$~mag, thus we did not overestimate
the variation.
In order to confirm the significance of the observed bump, we calculate
the variability parameter $C$ (\cite{Romero}) with the scale factor
$\Gamma$ (\cite{Howell88} and \cite{Cellone07a}).
According to \citet{Romero}, an object can be considered as variable (at the
99 \% confidence level) if their parameters
satisfies $C/{\Gamma}\ {\geq}\ 2.576$.
When the magnitudes of the object, comparison and check star were 12.6,
12.48 and 13.55, the $\Gamma$ was about 0.88, and the scaled variability
parameter $C/{\Gamma}$ around the bump ($t<17.68$ and $18.27<t$ in UT)
was 5.340. Thus the bump was not
caused by the effect of the magnitudes of the comparison and the check
star. 
Therefore, we conclude that the observed 15-min bump is a real one.

The $V-J$ color became bluer with the overall brightening trend. The
color at the end of the observation was
$0.025{\pm}0.011$~mag bluer than that at the start. Also at the 15-min
bump maximum, it became bluer, ${\Delta}(V-J)=-0.025{\pm}0.011$, and
this behavior is called as ``bluer-when-brighter''. Such a feature
of bluer-when-brighter was
also observed in the internight variation of one day time-scale in
S5~0716$+$714. In a long term trend of tens of days time-scale,
however, the feature of bluer-when-brighter was seldom seen
(\cite{Wagner} and \cite{Ghisellini}). This is the first time that
this feature was detected in such a short variability (15-min bump) in
optical and NIR bands. Our observation indicates that microvariability
has the same feature as the internight variability. It suggests
that the internight variability and the microvariability are caused by
the same mechanism, while the long term variation has a different one.

In our monitoring of S5~0716$+$714 in 2007, there are the other two days in
which we obtained long time-series data, that is, 3 Nov. (3.5 hours) and 13 
Nov. (2.5 hours). We investigate whether the frequency of
microvariability is high or low using the scaled variability parameter
$C/{\Gamma}$.
For our observations, the scaled variability parameters $C/{\Gamma}$ are 
1.22 and 1.04 on 3 and 13 Nov, respectively, in V band.
Based on this result, we conclude that the microvariability only
appeared on 20 Oct in our monitoring.

\subsection{Polarimetric observation}
\begin{figure}
  \begin{center}
 \FigureFile(80mm,55mm){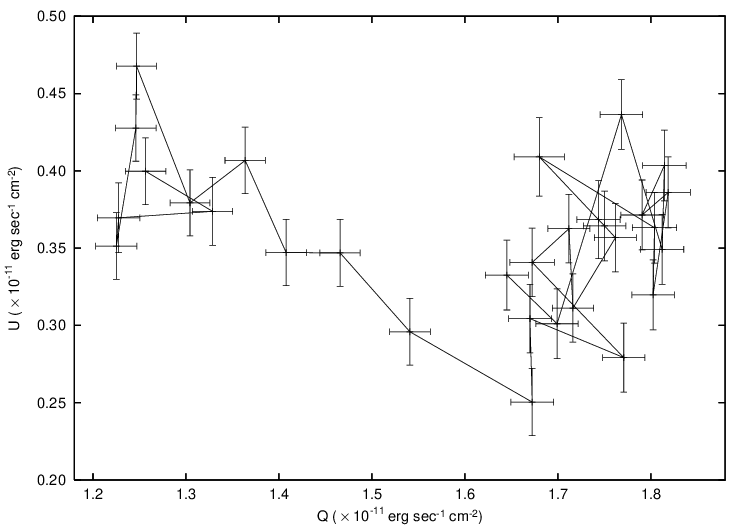}
  \end{center}
  \caption{%
    Stokes $Q$ and $U$ parameters in $V$ band during observations on 20
 Oct. 2007. The Stokes parameters have been rebinned to 4 min. The
 unit of the $Q$ and $U$ is ${\rm erg}\ {\rm sec}^{-1}\ {\rm cm}^{-2}$. }%
  \label{fig:qu}
\end{figure}

The polarization parameters were variable on 20 Oct, too. The maximum
polarization degree of the object was about 10.5 \% and minimum of that
was about 7.5 \%. Figure~2 shows the 
variation on the $Q-U$ plane. Time-series of the Stokes parameters, $Q$
and $U$, are depicted in figure~3, with the $V$-band light curve for
comparison. The temporal evolution of the Stokes parameters indicates
the presence of a variation feature on the $Q-U$ plane associated with
the bump. We calculated differential polarization vectors from the overall
trend. We assumed that there were two components in the
polarization vector; the first component was ${\mathbf{P}}_{\rm base}(t)$ 
associated with the overall brightening trend and the second one was 
${\mathbf{P}}_{\rm diff}(t)$ which was a deviation from 
${\mathbf{P}}_{\rm base}(t)$. They are written as,
\begin{eqnarray}
{\mathbf{P}}_{\rm obs}(t) = {\mathbf{P}}_{\rm base}(t) + {\mathbf{P}}_{\rm diff}(t) \\
{\mathbf{P}}_{\rm base}(t) = (Q_{\rm base}(t), U_{\rm base}(t)).
\end{eqnarray}
In our analysis, $Q_{\rm base}(t)$  and $U_{\rm base}(t)$ are approximated as
linear functions which are obtained by fitting the data outside of the bump 
($t<17.68$ and $18.27<t$ in UT), as shown in figure~3. We calculated the
differential polarized flux, 
$PF_{\rm diff}(t)=|{\mathbf{P}}_{\rm diff}(t)|$, and angle, 
$PA_{\rm diff}(t)=0.5arctan[U_{\rm diff}(t)/Q_{\rm diff}(t)]$,
which are shown in figure~4.
\begin{figure}
  \begin{center}
 \FigureFile(80mm,55mm){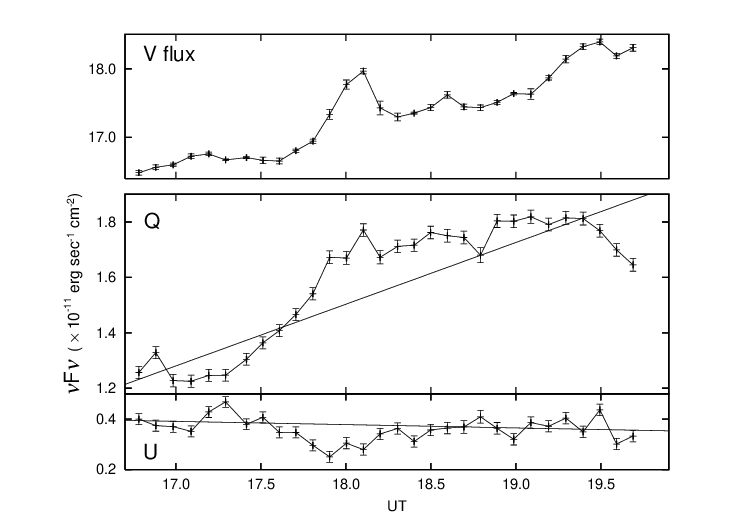}
  \end{center}
  \caption{%
 Top panel: The flux light curve in $V$ band. Middle and bottom panel:
 $Q$ and $U$ curves in the same band. Solid
 lines indicate the ${\mathbf{P}}_{\rm base}(t)$, which is defined in
 the text. All units of the vertical axis are 
 ${\rm erg}\ {\rm sec}^{-1}\ {\rm cm}^{-2}$. The flux was calculated with 1.98
 ${\times}10^{-5}{\rm erg}\ {\rm sec}^{-1}\ {\rm cm}^{-2}$ at 0 mag in
 $V$ band (\cite{Fukugita95}).} 
  \label{fig3}
\end{figure}
\begin{figure}
  \begin{center}
 \FigureFile(80mm,55mm){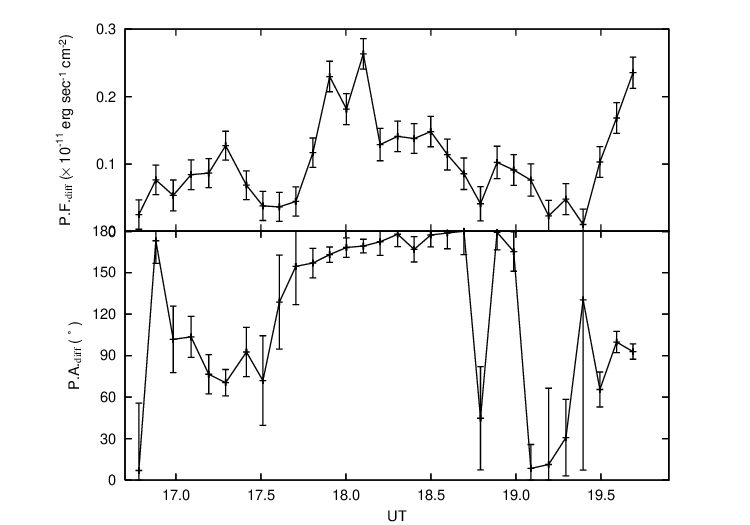}
  \end{center}
  \caption{%
   Time variation of the differential polarization vector from the
 overall brightening trend. The top figure shows the differential
 polarized flux 
 $PF_{\rm diff}(t)$ $({\rm erg}\ {\rm sec}^{-1}\ {\rm cm}^{-2})$. At
 around 18 UT, there is a peak which is associated with
 the bump. The bottom figure shows the differential polarization angle
 $PA_{\rm diff}(t)$ (degree).}%
  \label{fig4}
\end{figure}

If the bump component has no specific polarization vector, the 
$PF_{\rm diff}(t)$
and $PA_{\rm diff}(t)$ should be mainly constructed by a random noise. In
figure~4, however, $PF_{\rm diff}(t)$ is correlated with the
bump. The $PA_{\rm diff}(t)$ dramatically changed just before the
bump at 17.5 UT, and then, it kept constant during the bump.
Thus the bump had the intrinsic polarization vector whose polarization
angle was constant. This result is insensitive to the data period used
for the definition of the overall trend. Although the
$PA_{\rm diff}(t)$ apparently continues to be constant even
after the bump, it may be an artificial effect, due to the linear
fitting to the overall trend.

Figure~4 indicates another bump-like structure in $PF_{\rm diff}(t)$ at
19.5 UT. It can also be associated with a possible bump in the light
curve, as can be seen from figure~1 and 3. Thus, such a short bump may
frequently appear in the brightest state of S5~0716$+$714.

\section{Discussion}
The time-scales in the object frame were
$740{\pm}50$ and $480{\pm}30$ sec in the rising and decaying phases of
the bump. The size of the emitting region, $R$, can be estimated as a
light-crossing time, $R<c{\Delta}{\tau}$, where $c$ is the speed of
light. With the correction of the Doppler boosting, it is now,
$R<{\delta}c{\Delta}{\tau}$. A Schwarzschild radius in the black hole
mass $M$ is defined as $R_{\rm S}=2GM/c^{2}$, where $G$ is the
gravitational constant. If the object has typical values of the Doppler
factor ${\delta}{\approx}10$ and the black hole mass
$M=10^{9}M_{\odot}$, the emitting region $R$ is  constrained to be 
$R<(0.49{\pm}0.03){\cdot}R_{\rm S}$. In this case, thus, the variation
originated from a quite confined region whose size is smaller than
$R_{\rm S}$.
We now assume that the bump was caused by variations of the whole area
in the jet at a certain distance apart from the black hole. In this
situation, we can expect that the emitting region is larger than the
Schwarzschild radius, $R>R_{\rm S}$, except for a scenario with an
extreme re-confinement of the jet (\cite{ Sokolov}). In this case,
however, the emitting region is smaller than $R_{\rm S}$. In order to
satisfy $R>R_{\rm S}$, $M$ should be
\begin{eqnarray}
\frac{M}{\delta} < \frac{c^{3}{\Delta}{\tau}}{2G}. \label{mass-gamma}
\end{eqnarray}
Substituting the decaying time-scale of 480~sec to
equation~(\ref{mass-gamma}), we obtain  
${\delta}>(20{\pm}1){\times}(M/10^{9}M_{\odot})$. 
The bump could, thus, originate from the whole area of the jet if
S5~0716$+$714 has a relatively small black hole mass or a large Doppler
factor. 

The whole emitting region in blazars is, however, 
believed to be not so confined in ${\rm 1}{\cdot}R_{\rm S}$. If the emitting region is
5 or ${\rm 10}{\cdot}R_{\rm S}$, the relationship with $\delta$ and $M$ is about 
${\delta}>(100{\;}{\rm or}{\;}200){\times}(M/10^{9}M_{\odot})$.
In general, the optical emitting region is believed to be extended in sub-pc
scale in jets. If the emitting region is located at the sub-pc scale
from the black hole, the black hole mass (or Doppler factor) would be
extraordinarily small (or large), in order to explain the bump by the
variation of the whole emitting region.
Thus, an alternative idea is required. The idea is that the emission
region of the microvariability is a small and local area in the
whole area of the jet. This suggestion is consistent with the small
amplitude of the magnitude at the bump. 

In section~3, we showed that the 15-min bump has an intrinsic
polarization vector. We estimate the polarization degree of the bump
component. The maximum value of the differential polarized flux is
$(2.7{\pm}0.5){\times}10^{-12}$ ${\rm erg}\ {\rm sec}^{-1}\ {\rm cm}^{-2}$.
Using the total flux at the bump peak, $(1.02{\pm}0.02){\times}10^{-11}$ 
${\rm erg}\ {\rm sec}^{-1}\ {\rm cm}^{-2}$, the polarization degree of
the bump component is calculated to be $27{\pm}5\%$. On the other
hand, the observed polarization degree at the bump is
$9.8{\pm}0.5\%$. Therefore, the polarization degree of the bump
component is much larger than that of the emission component of the
overall brightening trend. The bump, hence, presumably originated
from the region where the magnetic field is more aligned than that in
the emitting region of the overall brightening trend. 

On the basis of both variability time-scale and the polarization
behavior, we suggest that the bump originated not from the whole jet,
but from the local region where the magnetic field is aligned.

\section{Conclusion}
We obtained multicolor photometric and polarimetric data of the shortest
time-scale variation in S5~0716$+$714 for the first time. Our findings
are summarized as below; first, the object became blue with
${\Delta}(V-J)=-0.025$ during the bump. Second, the bump
component has a specific polarization vector with the large polarization
degree of $27\%$, which is distinct from the overall brightening
trend. It can be suggested that the emitting region of the
microvariability in optical band is a small and local area compared with the
whole optical emitting region in the jet. 
\\
\\
This work was partly supported by a Grant-in-Aid from the Ministry of
Education, Culture, Sports, Science, and Technology of Japan (19740104).

\end{document}